\documentclass[11pt,a4p]{article}

\usepackage{epsfig}
\usepackage{cite}

\newcommand{\mulc}{\multicolumn}
\newcommand{\ra}{\rightarrow}
\newcommand{\ee}{{\rm e^+e^-}}

\newcommand{\qq}{{\rm q \bar q}}
\newcommand{\mumu}{\mu^+\mu^-}

\newcommand{\ellell}{\ell^+ \ell^-}

\newcommand{\Z}{{\rm Z}}

\newcommand{\hee}{{\rm h_{ee}}}
\newcommand{\heesq}{{\rm h_{ee}^{2}}}

\newcommand{\hll}{{\rm h_{\ell\ell}}}
\newcommand{\hpmpm}{{\rm H}^{\pm\pm}}

\begin{document}

%=======================================================================
%=======================================================================

\begin{titlepage}
\begin{center}{\large
    EUROPEAN ORGANIZATION FOR NUCLEAR RESEARCH
    }
\end{center}
\bigskip
  \begin{flushright}
    CERN-EP/2003-041\\
    July 14, 2003 \\
  \end{flushright}
  \bigskip\bigskip\bigskip\bigskip\bigskip
  \begin{center}{\huge\bf\boldmath
      Search for the Single Production of Doubly-Charged
      Higgs Bosons and Constraints on their Couplings
      from Bhabha Scattering
      }
  \end{center}
  
  \bigskip\bigskip

\begin{center}{\Large
  The OPAL Collaboration
  }
\end{center}

  \bigskip\bigskip
  
\begin{abstract}
   A search for the single production of doubly-charged Higgs bosons
   is performed using $\ee$ collision data collected by the OPAL 
   experiment at centre-of-mass energies between 189~GeV and 209~GeV.
   No evidence for the existence of $\hpmpm$ is observed.
   Upper limits are derived on $\hee$, the Yukawa coupling of the 
   $\hpmpm$ to like-signed electron pairs.
   A 95\% confidence level upper limit of $\hee<$~0.071 is inferred
   for $M(\hpmpm)<$~160~GeV assuming that the sum of the branching 
   fractions of the $\hpmpm$ to all lepton flavour combinations is 100\%.
   Additionally, indirect constraints on $\hee$ from Bhabha scattering at
   centre-of-mass energies between 183~GeV and 209~GeV, where the 
   $\hpmpm$ would contribute via $t$-channel exchange, are derived for 
   $M(\hpmpm)<$~2~TeV.
   These are the first results both from a single production search
   and on constraints from Bhabha scattering reported from LEP.
\end{abstract}

  \bigskip\bigskip
  \bigskip\bigskip
  
  \begin{center} Submitted to Phys.Lett.B \end{center}

\end{titlepage}

\begin{center}{\Large        The OPAL Collaboration
}\end{center}\bigskip
\begin{center}{
%begin authorlist PLEASE DO NOT DELETE THIS COMMENT
G.\thinspace Abbiendi$^{  2}$,
C.\thinspace Ainsley$^{  5}$,
P.F.\thinspace {\AA}kesson$^{  3}$,
G.\thinspace Alexander$^{ 22}$,
J.\thinspace Allison$^{ 16}$,
P.\thinspace Amaral$^{  9}$, 
G.\thinspace Anagnostou$^{  1}$,
K.J.\thinspace Anderson$^{  9}$,
S.\thinspace Arcelli$^{  2}$,
S.\thinspace Asai$^{ 23}$
% ...
%end authorlist PLEASE DO NOT DELETE THIS COMMENT
}\end{center}\bigskip
\bigskip
%begin institutes
$^{  1}$School of Physics and Astronomy, University of Birmingham,
Birmingham B15 2TT, UK
\newline
$^{  2}$Dipartimento di Fisica dell' Universit\`a di Bologna and INFN,
I-40126 Bologna, Italy
\newline
$^{  3}$Physikalisches Institut, Universit\"at Bonn,
D-53115 Bonn, Germany
\newline
% ...
$^{  5}$Cavendish Laboratory, Cambridge CB3 0HE, UK
\newline
% ...
$^{  9}$Enrico Fermi Institute and Department of Physics,
University of Chicago, Chicago IL 60637, USA
\newline
% ...
$^{ 16}$Department of Physics, Schuster Laboratory, The University,
Manchester M13 9PL, UK
\newline
% ...
$^{ 22}$Department of Physics and Astronomy, Tel Aviv University,
Tel Aviv 69978, Israel
\newline
$^{ 23}$International Centre for Elementary Particle Physics and
Department of Physics, University of Tokyo, Tokyo 113-0033, and
Kobe University, Kobe 657-8501, Japan
\newline
% ...
%end notes

%=======================================================================

\section{Introduction}

\normalsize

\label{s:intro}

Some theories beyond the Standard Model predict the existence of
doubly-charged Higgs bosons, $\hpmpm$, including in Left-Right
Symmetric models \cite{ref:lrmodels}, Higgs Triplet models \cite{ref:triplet}, 
and little Higgs models~\cite{ref:littlehiggs}.  It has been particularly 
emphasized that a see-saw mechanism used to obtain light neutrinos in a 
model with heavy right-handed neutrinos can lead to a doubly-charged Higgs
boson with a mass accessible to current and future colliders
\cite{ref:seesaw}.

A review of experimental constraints on doubly-charged Higgs bosons is 
presented in \cite{ref:swartz}. The pair production of doubly-charged Higgs 
bosons has been considered in a previous OPAL publication \cite{ref:pr349}, 
where masses less than 98.5~GeV are excluded for doubly-charged Higgs
bosons in Left-Right Symmetric models. DELPHI has obtained a limit of
97.3~GeV, independent of the lifetime of the $\hpmpm$~\cite{Abdallah:2002qj}.

\begin{figure}[h]
  \begin{center}
      \mbox{
        \epsfig{file=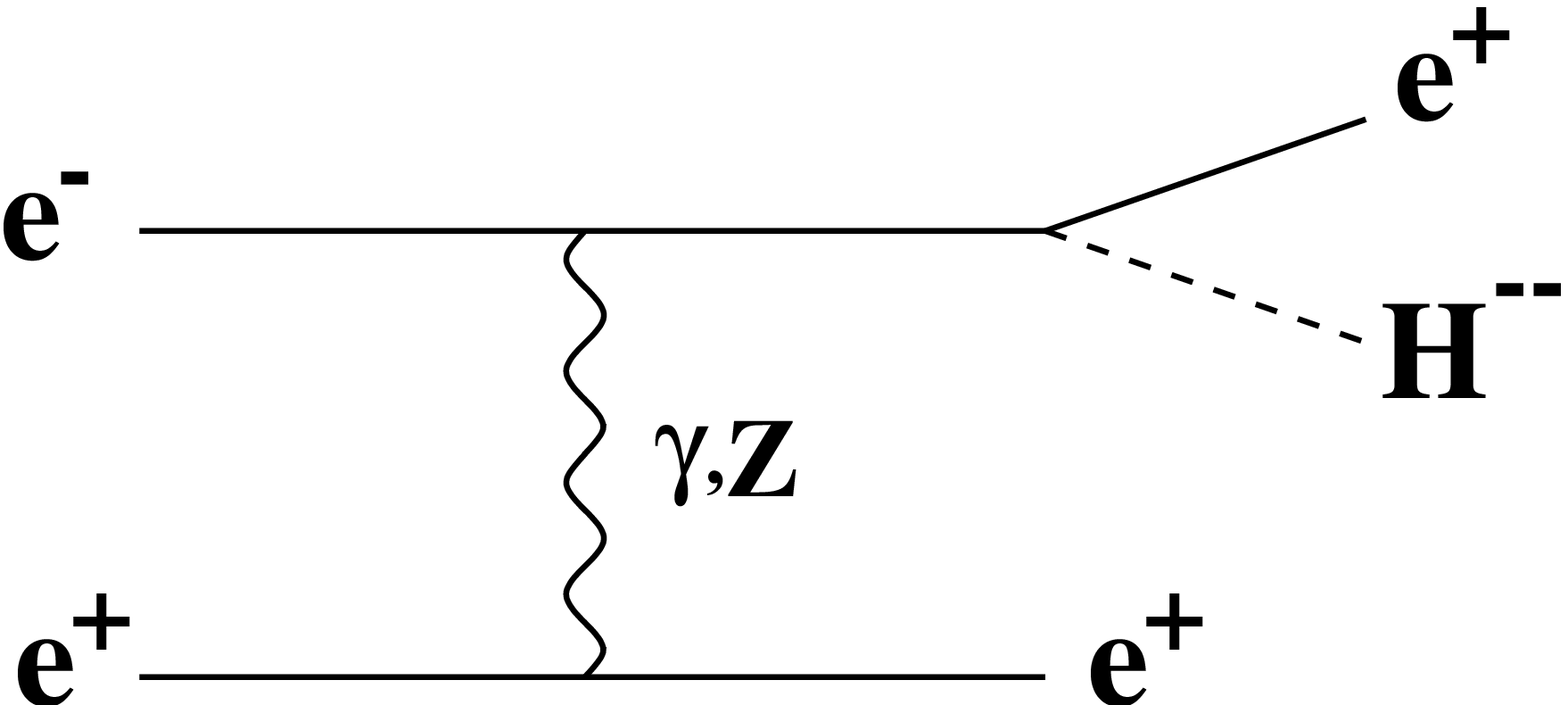,width=0.35\textwidth
          }
        \epsfig{file=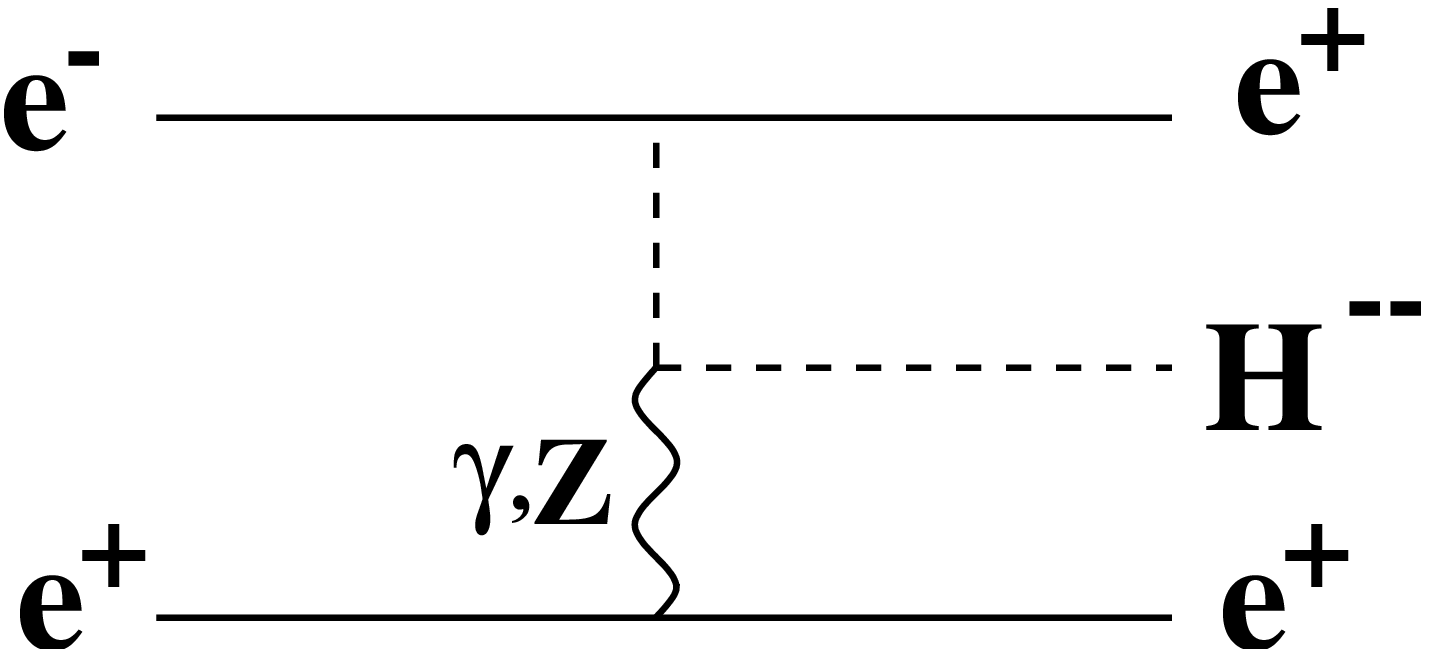,width=0.3\textwidth
          }
        \epsfig{file=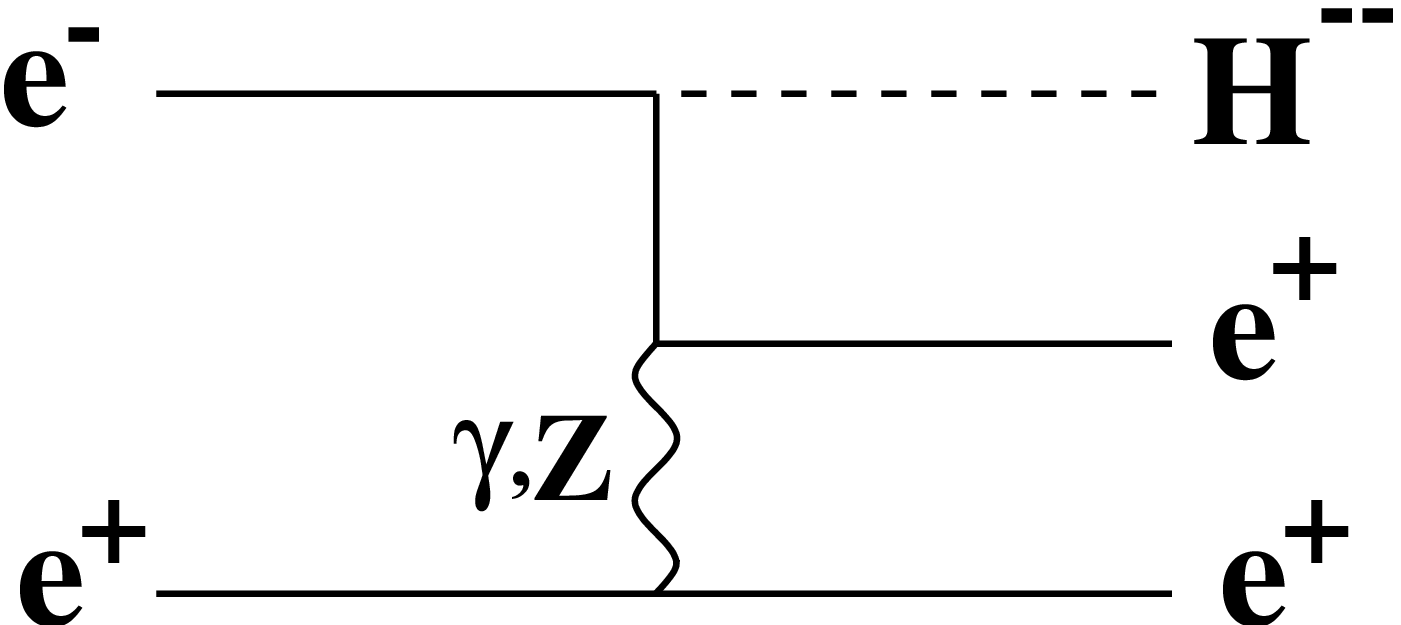,width=0.3\textwidth
          }
        }
  \end{center}
\caption{\label{f:direct}
  \protect{
    \sl Feynman diagrams contributing to the single production
    of $\rm H^{--}$ bosons in $\ee$ collisions.  The
    three additional diagrams with ``crossed'' $\rm e^+$ lines
    are not shown.
  }
}
\end{figure}

It has been noted that doubly-charged Higgs bosons may be singly
produced in $\rm e\gamma$ collisions, including in $\ee$ collisions
where the $\gamma$ is obtained from radiation from the other beam
particle \cite{Barenboim:1996pt, ref:godkal}.  The diagrams for the
direct production are shown in Figure~\ref{f:direct}.

Doubly-charged Higgs bosons would decay into like-signed lepton or
vector boson pairs, or to a W boson and a singly-charged Higgs
boson.  For masses less than twice the W boson mass, they would decay
predominantly into like-signed leptons. Furthermore, in most models
the WW branching fraction is negligible even for larger masses
\cite{ref:godkal}, therefore the dominant decay mode, even for masses
larger than twice the W boson mass, is the decay to like-signed
leptons.
Since the $\hpmpm$ naturally violates lepton number conservation, it
can have mixed lepton flavour decay modes.  Additionally, the
Yukawa coupling of the $\hpmpm$ to the charged leptons h$_{\ell\ell}$
is model dependent, and is not generally determined directly by the
lepton mass, so decays to all lepton flavour combinations need to be
considered. It should be particularly noted that mixed lepton flavour
decays are severely constrained by rare decay searches such as
$\mu^+\ra\rm e^+e^+e^-$ and $\mu\ra\rm e\gamma$.

In this paper, we search for the single production of doubly-charged
Higgs bosons, assuming the decays $\hpmpm\ra\ell^\pm\ell^{\prime\pm}$
using 600.7~pb${^{-1}}$ of $\ee$ collision data with centre-of-mass
energies $\sqrt{s}=$~189--209~GeV collected by the OPAL detector.
Since the production cross-section depends only on $\hee$, the Yukawa
coupling of the $\hpmpm$ to like-signed electron pairs, the search is
sensitive to this quantity.

We assume that the decay of a doubly-charged Higgs boson into a
W boson and a singly-charged Higgs boson is negligible. We consider
an $\hpmpm$ which couples to right-handed particles, but the results 
of the direct search quoted here are also valid for an $\hpmpm$ which 
couples only to left-handed particles~\cite{ref:godkal}. All
lepton flavour combinations are considered in the $\hpmpm$ decay ($\rm
ee$, $\mu\mu$, $\tau\tau$, $\rm e\mu$, $\rm e\tau$, $\mu\tau$).  The
lifetime of the $\hpmpm$ can be important, and in particular is
non-negligible for $\hll<10^{-7}$; however, our search is not
sensitive to such small Yukawa couplings.

\bigskip

\begin{figure}[ht]
  \begin{center}
      \mbox{
        \epsfig{file=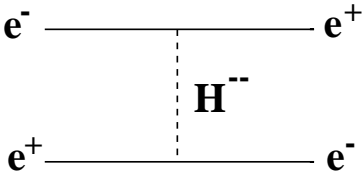,width=0.3\textwidth
          }
        }
  \end{center}
\caption{\label{f:indirect}
  \protect{
    \sl Feynman diagram contributing to the process $\ee\ra\ee$
    due to doubly-charged Higgs boson $t$-channel exchange.
  }
}
\end{figure}

A doubly-charged Higgs boson would also affect the Bhabha scattering
cross-section via the $t$-channel exchange diagram shown in
Figure~\ref{f:indirect}, causing a change in rate and in the observed
angular distribution of the outgoing electron.  Constraints have been
derived for this process using data from lower energy colliders
\cite{ref:swartz}, but not previously from LEP.

In addition to the direct search results introduced above, we also
derive indirect constraints on $\hee$, the Yukawa coupling of $\hpmpm$
to electrons, using the differential cross-section of wide-angle
Bhabha scattering measured by OPAL in 688.4~pb${^{-1}}$ of data
collected at $\sqrt{s}=$~183--209~GeV.

%=======================================================================

\section{OPAL Detector}

\label{s:detector}

The OPAL detector is described in detail in~\cite{ref:OPAL-detector}.
It is a multipurpose apparatus with almost complete solid angle coverage.
The central detector consists of a silicon micro-strip detector and
a system of gas-filled tracking chambers in a 0.435\,T solenoidal magnetic 
field which is parallel to the beam axis. A lead-glass electromagnetic 
calorimeter with a presampler surrounds the central detector.
In combination with the forward calorimeters, the forward scintillating-tile 
counters, and the silicon-tungsten luminometer, a geometrical acceptance 
is provided down to 25\,mrad from the beam direction.
The silicon-tungsten luminometer measures the integrated luminosity using 
small-angle Bhabha scattering events. The magnet return yoke is instrumented 
for hadron calorimetry, and is surrounded by several layers of muon chambers.

%=======================================================================

\section{Direct Search}

\label{s:direct}

%---------------------------------------------------------------------------

\subsection{Data Samples and Event Simulation}

\label{ss:data}

The data samples are summarised in Table~\ref{t:data}.

\begin{table}[ht]
\centering
  \begin{tabular}{c|c|c}
\hline
    $E_{\rm cm}$    & $\langle E_{\rm cm}\rangle$ & $\int \cal{L}$ \\
      (GeV)         &   (GeV)        &    ($\rm pb^{-1}$) \\\hline
     188   -- 190   &  188.6         & 175.0 \\
     190   -- 194   &  191.6         &  28.9\\
     194   -- 198   &  195.5         &  74.8\\
     198   -- 201   &  199.5         &  78.1\\
     201   -- 203   &  201.7         &  38.2\\
     203   -- 206   &  205.0         &  79.4\\
     206   -- 209   &  206.6         & 126.1\\\hline
     188 -- 209     &  197.7         &  600.7 \\\hline
  \end{tabular}
  \caption{\label{t:data}
    \sl Data samples used in the direct search analysis.}
\end{table}

The process $\rm\ee\ra e^\mp e^\mp\hpmpm$ is simulated with the 
PYTHIA6.150~\cite{ref:PYTHIA} event generator.
In the simulation, the Equivalent Photon Approximation (EPA) is used 
to give an effective flux of photons originating from the electrons 
or positrons. The upper limit of the virtuality $Q^2$ of the photon 
is given by the scale of the hard scattering process\footnote{$Q^2$ 
is the negative squared four-momentum transfer.}.
The process $\rm e^\pm\gamma\ra e^{\mp}\hpmpm$ is simulated in a 
Left-Right Symmetric model for an $\hpmpm$ which couples to 
right-handed particles using the calculations 
from \cite{Barenboim:1996pt}. The contribution from $\Z$-exchange
is negligible.
In order to obtain the full signal cross-section, a cut which PYTHIA
applies by default at a minimum of 1~GeV on the transverse momentum of the 
lepton which radiates the $\hpmpm$ is explicitly switched off.
The cross-section and the angular distribution are checked with the 
calculations of \cite{ref:godkal}, using COMPHEP \cite{ref:comphep}. 

Separate samples are simulated with the 6 different decay modes 
($\rm ee$, $\mu\mu$, $\tau\tau$, $\rm e\mu$, $\rm e\tau$, $\mu\tau$).
Samples of 500 events each are generated for each of the average 
centre-of-mass energies listed in Table~\ref{t:data} for $\hpmpm$ 
masses in 5~GeV steps from 90--200 GeV.  
For masses larger than twice the W boson mass the decay $\rm \hpmpm \ra W^{\pm}W^{\pm}$
is kinematically allowed. Its partial width, however, is negligible in
most models~\cite{ref:godkal}.
In this paper, the branching fraction BR($\rm \hpmpm \ra W^{\pm}W^{\pm}$) 
is assumed to be zero.

The dominant Standard Model backgrounds in this analysis are
from the four-fermion processes
$\ee\ra\ell^+\ell^-\ell^{\prime +}\ell^{\prime -}$, including
events from the so-called ``multi-peripheral'' diagrams
$\ee\ra\ee\gamma^{(*)}\gamma^{(*)}\ra\ee\ell^+\ell^-$, and lepton pairs,
$\ee\ra\ell^+\ell^-$. Four-fermion processes except 
$\ee\ra\ee\ell^+\ell^-$ ($\ell={\rm e}, \mu,\tau$) and $\ee\ra\ee\qq$ 
are simulated with the KORALW event generator~\cite{ref:koralw}.
The non-multi-peripheral part of the processes $\ee\ra\ee\ell^+\ell^-$ 
and $\ee\ra\ee\qq$ is simulated with grc4f2.1\cite{ref:grc4f}.
The multi-peripheral diagrams are simulated with the dedicated two-photon 
event generators Vermaseren~\cite{ref:Vermaseren} for
$\ee\ra\ee\gamma^{(*)}\gamma^{(*)}\ra\ee\ee$ and
BDK~\cite{ref:bdk} for
$\ee\ra\ee\gamma^{(*)}\gamma^{(*)}\ra\ee\mu^+\mu^-$ and
$\ee\ra\ee\gamma^{(*)}\gamma^{(*)}\ra\ee\tau^+\tau^-$.
The Monte Carlo generators
PHOJET~\cite{ref:PHOJET} (for $Q^2 < 4.5$~GeV$^2$) and 
HERWIG~\cite{ref:HERWIG}
(for $Q^2 \geq 4.5$~GeV$^2$) are used to simulate hadronic events 
from two-photon processes.
Lepton pairs are simulated using
the KK2f~\cite{ref:kk2f} generator for
$\tau^+ \tau^- (\gamma)$ and $\mumu (\gamma)$ events and
NUNUGPV~\cite{ref:nunugpv} for 
$\nu\bar\nu\gamma(\gamma)$.
Bhabha scattering is simulated with
BHWIDE~\cite{ref:BHWIDE} (when both the electron and
positron scatter at least 12.5$^\circ$ from the beam axis)
and TEEGG~\cite{ref:teegg} (for the remaining phase space).

Multihadronic events, $\qq (\gamma)$, are simulated using
KK2f~\cite{ref:kk2f}.
RADCOR~\cite{ref:radcor} is used to simulate multi-photon events
from QED processes.
They make a negligible contribution to the background.

Generated signal and background events are processed
through the full simulation of the OPAL detector~\cite{ref:GOPAL}
and the same event analysis chain was applied to the simulated events
as to the data.

%---------------------------------------------------------------------------

\subsection{Analysis}

\label{ss:analysis}

The signal final state consists of four charged leptons. Two like-sign
leptons originate from the $\hpmpm$ decay and are expected to be visible
in the detector in most cases. The electron or positron which originates 
from the ee$\gamma$ vertex (see Fig.~\ref{f:direct}) in general escapes 
through the beampipe. The electron or positron which originates from the 
ee$\hpmpm$ vertex is also forward peaked; however, it enters the detector 
in a significant fraction of signal events. 
The analysis is therefore divided into a two-lepton and a three-lepton analysis.
The final states in the three-lepton case contain three leptons visible
in the detector, two of them have the same sign and could originate from the
decay of a doubly-charged Higgs boson. 
In the two-lepton case, two like-signed leptons are required, as expected in the
decay of a doubly-charged Higgs boson.

Leptons are identified as low multiplicity jets. Jets are reconstructed 
from charged particle tracks and energy deposits (clusters) in the
electromagnetic and hadron calorimeters.
Tracks and clusters are defined to be of ``good'' quality using
the requirements of \cite{ref:MT}.
After the jet reconstruction, double-counting of energy between tracks and 
calorimeter clusters is corrected by reducing the calorimeter cluster energy
by the expected energy deposition from associated charged tracks \cite{ref:MT}, 
including particle identification information.

\begin{figure}[b]
  \begin{center}
      \mbox{\epsfig{file=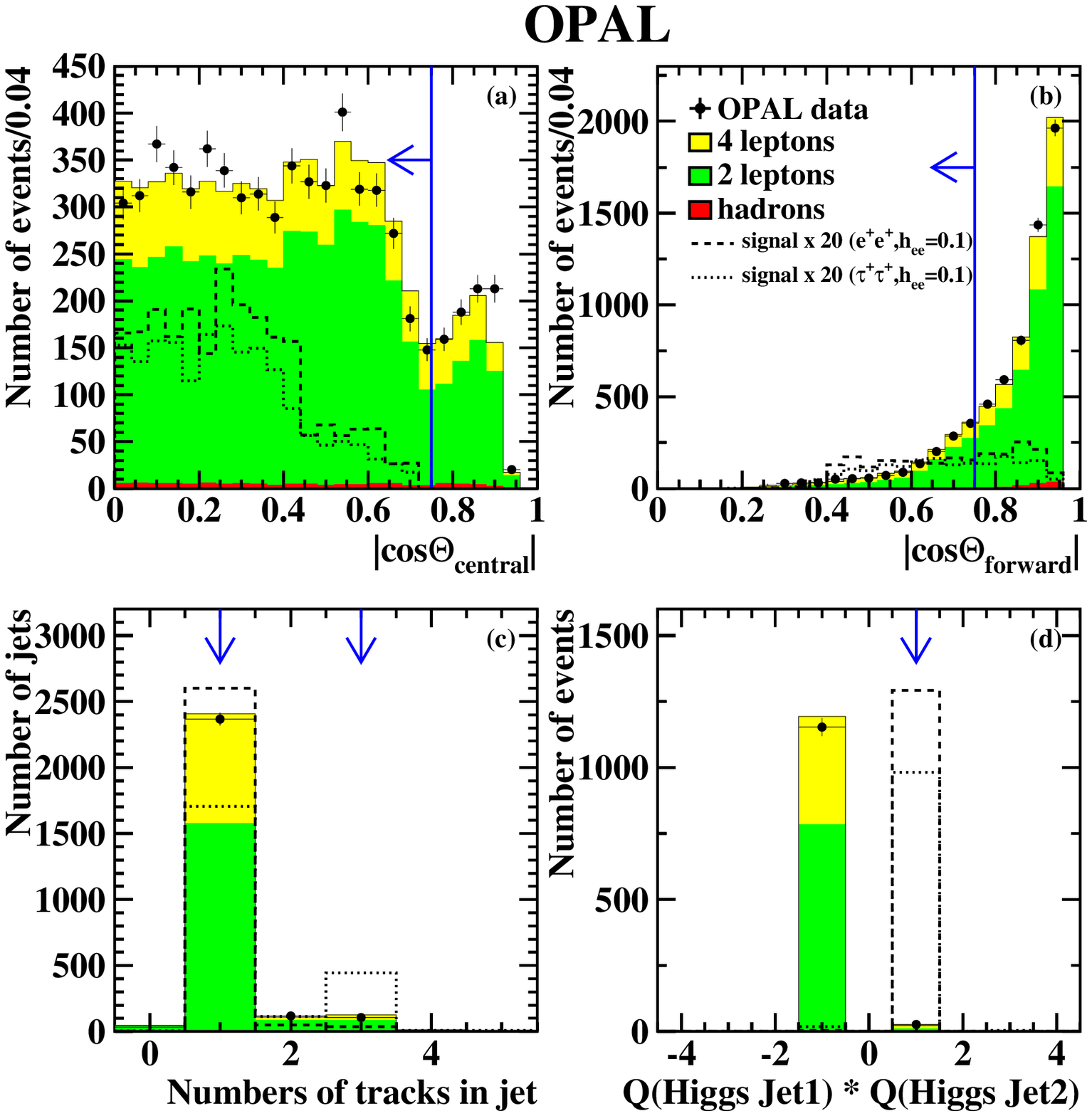,width=\textwidth
          }}
  \end{center}
\caption{\label{f:cuts2l}
  \protect{\sl
  Examples of some of the quantities used
  in the two-lepton analysis selection shown immediately before
  the corresponding cut is applied (see Section~\ref{ss:analysis}).
  The absolute values of the cosines of the polar angle
  of the more central and the more forward Higgs boson candidate jets
  are shown in (a) and (b), the number of charged tracks in each of 
  the two $\hpmpm$ candidate jets in (c), and the product of the 
  reconstructed charges of the two $\hpmpm$ candidate jets in (d).
  The points with error bars indicate the OPAL data and  the shaded regions
  indicate the background expectation. Note that ``hadrons'' includes both
  $\qq (\gamma)$ and hadronic events from all 4-fermion processes.
  Two example signal expectations for a 130 GeV doubly-charged Higgs boson 
  are also shown normalised to a cross-section corresponding to  $\hee=0.1$ 
  scaled by a factor 20, assuming either a 100\% $\hpmpm\ra\rm ee$ branching
  ratio (dashed line) or a 100\% $\hpmpm\ra\tau\tau$ branching ratio 
  (dotted line). The cut requirements are indicated by the arrows.
  }
}
\end{figure}

No explicit electron or muon identification is required, since it is found that
the jet-based analysis technique retains high efficiency while
reducing the background to an acceptable level.
The same analysis is used to search for all 6 possible lepton flavour combinations, 
and the results are valid for all leptonic decay modes of the $\hpmpm$.
The final background is dominated by Standard Model processes containing four 
charged leptons.
The analysis cuts are listed below. The cut values of the two-lepton and 
three-lepton analyses differ slightly.
\newline
The requirements for the two-lepton analysis are:
\begin{itemize}
  \item[(2.1)] The preselection requires low multiplicity events
    \cite{ref:ll}.
    The events are additionally required to have at least two and less
    than nine charged tracks. The sum of charged tracks and clusters
    in the electromagnetic calorimeter must be less than 16.
    Tracks and clusters are formed into jets using a cone algorithm~\cite{ref:cone} 
    with a half-angle of 20 degrees and a minimum jet energy of 2.5~GeV,
    and it is required that there be exactly two jets with polar 
    angles\footnote{ OPAL uses a right-handed coordinate system where the $+z$ 
    direction is along the electron beam and where $+x$ points to the 
    centre of the LEP ring. The polar angle $\theta$ is defined with respect 
    to the $+z$ direction and the azimuthal angle $\phi$ with respect to the 
    $+x$ direction. The centre of the $\ee $ collision region defines the 
    origin of the coordinate system.}
    satisfying $|\cos\theta|<$~0.95, and which are not precisely back-to-back
    (within 5$^\circ$). Finally, the sum of the energies of the two jets 
    reconstructed in the event must be greater than 20\% of $\sqrt{s}$.
  \item[(2.2)] Ordering the jet energies by their magnitude
             ($E_{\rm jet1}>E_{\rm jet2}$),
             the following requirements are made:
     \begin{itemize}
       \item[a)] $E_{\rm jet1} > 0.1 \sqrt{s}$;
       \item[b)] $E_{\rm jet2} > 0.05 \sqrt{s}$;
       \item[c)] $E_{\rm jet1} <  0.995 E_{\rm beam}$;
       \item[d)] $E_{\rm jet1}+E_{\rm jet2} <  0.95 \sqrt{s}$.
     \end{itemize}
   \item[(2.3)] The invariant mass $M_{\rm inv}$ of the two jets
     must satisfy $M_{\rm inv}>$~40~GeV.
     Typical mass resolutions are about 4~GeV for ee and 10~GeV for $\mu\mu$.
     No mass reconstruction is possible for $\tau\tau$, 
     due to the undetected neutrinos.
   \item[(2.4)] Bhabha scattering is rejected by requiring that the
     acollinearity angle, $\phi_{\rm acol}$, satisfies $\phi_{\rm acol}>25^\circ$.
     The angle $\phi_{\rm acol}$ is defined to be 180$^\circ$ minus
     the opening angle of the two jets.
   \item[(2.5)] The polar angle of each jet must satisfy $|\cos\theta|<0.75$.  
     The $\hpmpm$ candidate jet polar angles are plotted in 
     Figures~\ref{f:cuts2l}(a) and (b) after cuts (2.1)--(2.4).
   \item[(2.6)]  Each jet associated to the $\hpmpm$ must have either one or 
     three charged tracks. The number of charged tracks is plotted in
     Figure~\ref{f:cuts2l}(c) after cuts (2.1)--(2.5).
   \item[(2.7)] Defining the sum of the track charges within
     each jet as the ``jet charge'', the product of the charges
     of the two jets must be equal to $+1$.
     This value is plotted in Figure~\ref{f:cuts2l}(d)
     after cuts (2.1)--(2.6).
\end{itemize}

\begin{figure}[b]
  \begin{center}
      \mbox{\epsfig{file=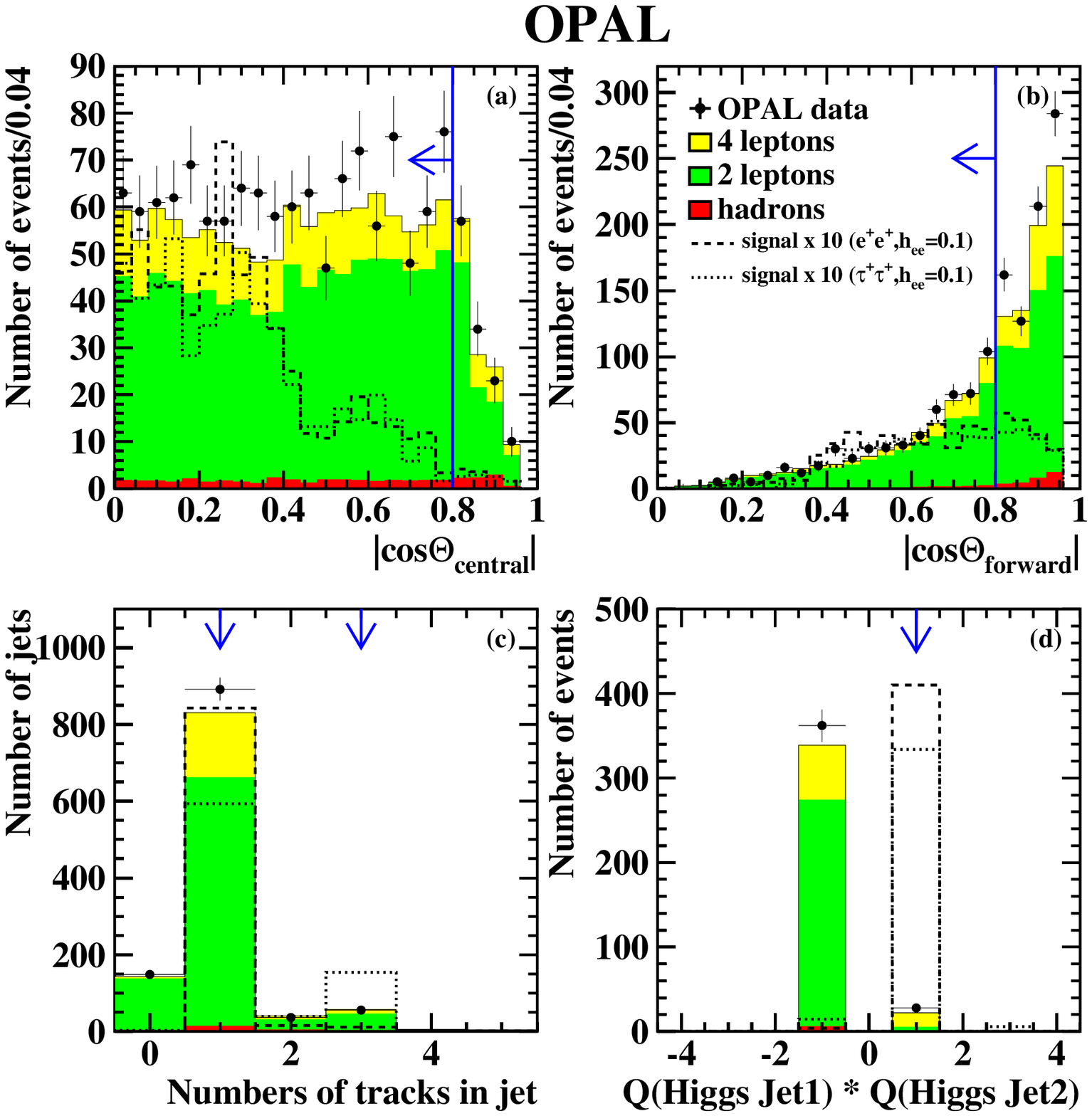,width=\textwidth
          }}
  \end{center}
\caption{\label{f:cuts3l}
  \protect{\sl
  Examples of some of the quantities used in the three-lepton analysis 
  selection shown immediately before the corresponding cut is applied
  (see Section~\ref{ss:analysis}).
  The absolute values of the cosines of the polar angle of the more 
  central and the more forward Higgs boson candidate jets
  are shown in (a) and (b), the number of charged tracks in each of 
  the two $\hpmpm$ candidate jets in (c), and the product of the 
  reconstructed charges of the two $\hpmpm$ candidate jets in (d).
  The points with error bars indicate the OPAL data and  the shaded regions
  indicate the background expectation. Note that ``hadrons'' includes both
  $\qq (\gamma)$ and hadronic events from all 4-fermion processes.
  Two example signal expectations for a 130 GeV doubly-charged Higgs 
  boson are also shown normalised to a cross-section corresponding to 
  $\hee=0.1$ scaled by a factor 10, assuming either a 100\% $\hpmpm\ra\rm ee$ 
  branching ratio (dashed line) or a 100\% $\hpmpm\ra\tau\tau$
  branching ratio (dotted line). The cut requirements are indicated by the arrows.
  }
}
\end{figure}

\vspace{1.0cm}
\noindent The requirements for the three-lepton analysis are:

\begin{itemize}
  \item[(3.1)] The preselection is identical to that in cut (2.1) except
    that exactly three reconstructed jets are required. The two
    jets which have the highest reconstructed mass, as described in cut 
    (3.3), have to satisfy $|\cos\theta|<$~0.95 and must not be precisely 
    back-to-back (within 5$^\circ$). There is no $|\cos\theta|$ requirement 
    for the third jet. Finally, the sum of the energies of the three jets 
    reconstructed in the event must be greater than 20\% of $\sqrt{s}$.
  \item[(3.2)] Ordering the measured jet energies by their magnitude
             ($E_{\rm jet1}>E_{\rm jet2}>E_{\rm jet3}$),
             the following requirements are made:
     \begin{itemize}
       \item[a)] $E_{\rm jet1} > 0.1 \sqrt{s}$;
       \item[b)] $E_{\rm jet2} > 0.05 \sqrt{s}$;
       \item[c)] $E_{\rm jet3} > 0.025 \sqrt{s}$ or it must
          contain at least one good charged track;
       \item[d)] $E_{\rm jet1} <  0.995 E_{\rm beam}$;
       \item[e)] $E_{\rm jet1}+E_{\rm jet2}+E_{\rm jet3} <  0.95 \sqrt{s}$.
     \end{itemize}
   \item[(3.3)]  
     The jet energies are determined assuming that the measured jet direction 
     is the same as the initial lepton direction for each of the reconstructed 
     jets and that the missing electron or positron is recoiling along the beam 
     axis. Using energy and momentum conservation to give four constraint
     equations, the four jet energies can be inferred (the lepton masses are 
     neglected). Using this improved determination of the jet energies, the 
     invariant masses are calculated for the three possible di-jet systems that
     can be constructed from the observed jets, and the two jets having the largest 
     di-jet mass are considered as the $\hpmpm$ candidate jets with a 
     ``reconstructed Higgs boson mass'' $M_{\rm rec}$.
     The loss due to this assumption is negligible for $\hpmpm$ masses
     above 110~GeV, and is taken into account in the signal efficiency
     calculation.
     Since this search concentrates on the region above the mass limit from pair 
     creation, it is further required that $M_{\rm rec}$ satisfy $M_{\rm rec}>$~80~GeV.
     Typical mass resolutions are about 1~GeV for ee and $\mu\mu$ modes, and about 
     4~GeV for $\tau\tau$ decays. Note that in the latter case, no mass reconstruction 
     from the jet energies would have been possible without this procedure, 
     due to the undetected neutrinos.
   \item[(3.4)] Bhabha scattering is rejected by requiring that the acollinearity angle 
     between the two $\hpmpm$ candidate jets satisfies $\phi_{\rm acol}>15^\circ$.
   \item[(3.5)] The polar angle of each jet associated to the $\hpmpm$ must satisfy 
     $|\cos\theta|<0.80$. The $\hpmpm$ candidate jet polar angles are plotted in 
     Figures~\ref{f:cuts3l}(a) and (b) after cuts (3.1)--(3.4).
   \item[(3.6)]  Each jet associated to the $\hpmpm$ must have either one or three 
     charged tracks. The number of charged tracks is plotted in
     Figure~\ref{f:cuts3l}(c) after cuts (3.1)--(3.5).
   \item[(3.7)] Defining the sum of the track charges within each jet as the 
     ``jet charge'', the product of the charges of the two jets associated with the 
     $\hpmpm$ must be equal to $+1$. This value is plotted in Figure~\ref{f:cuts3l}(d)
     after cuts (3.1)--(3.6).
\end{itemize}

The results are summarised in Table~\ref{t:results}. The numbers of observed and expected 
events agree well after each cut in both analyses.

\begin{table}[ht!] 
  \label{t:2-lepton analysis}
  \centering
  \begin{tabular}{l|rr|rrrrr|rrr}
    \multicolumn{11}{l}{\bf Two-lepton analysis}\\
    \hline
    Cut  & Data & Total     & $\ellell$ & 4-$\ell$ & `$\gamma\gamma$'&   $\qq$   & `$\gamma\gamma$' & \mulc{3}{c}{Efficiency [\%]}  \\
         &      & Bkg.      &           &          & $\rm ee\ell\ell$&           &   $\rm eeqq$     & \mulc{3}{c}{}   \\
         &      &           &           &          &                 &           &                  &   ee   & $\mu\mu$ & $\tau\tau$ \\ \hline
    (2.1)  & 19612& 17659.3   &  13776.9  &  1249.6  &     2249.7      &     173.3 &    209.9         & 45.7   &  45.9    &   41.5   \\
    (2.2)  & 15168& 14731.3   &  11381.3  &  1118.7  &     1971.5      &     158.1 &    101.8         & 44.3   &  39.1    &   36.4    \\
    (2.3)  & 13455& 13002.6   &  10855.6  &   988.1  &     1026.5      &     120.8 &     11.6         & 44.3   &  39.0    &   35.0    \\
    (2.4)  &  6681&  6685.9   &   5025.5  &   774.0  &      777.5      &     100.1 &      8.7         & 41.0   &  36.3    &   32.6    \\
    (2.5)  &  1318&  1353.4   &    890.6  &   325.9  &      124.3      &      12.5 &      0.1         & 23.8   &  24.3    &   20.3    \\
    (2.6)  &  1181&  1216.2   &    792.6  &   299.5  &      121.4      &       2.7 &      0.0         & 23.0   &  23.9    &   17.9    \\\hline
    (2.7)  &    27&    22.1   &     10.4  &     2.7  &        8.5      &       0.5 &      0.0         & 22.9   &  23.8    &   17.5    \\
         &      & $\pm$1.7  & $\pm$1.3  &  $\pm$0.2&  $\pm$1.1       &   $\pm$0.1&   $\pm$0.0     &$\pm$1.9  &$\pm$2.0  &$\pm$2.0  \\\hline
         &      &           &           &          &                 &           &                  & (64.6)   &  (67.3)    &   (49.1)   \\\hline
\multicolumn{11}{c}{}   \\
  \end{tabular}

  \label{t:3-lepton analysis}
  \centering
  \begin{tabular}{l|rr|rrrrr|rrr}
    \multicolumn{11}{l}{\bf Three-lepton analysis}\\
    \hline
    Cut  & Data & Total     & $\ellell$ & 4-$\ell$ & `$\gamma\gamma$'&   $\qq$   & `$\gamma\gamma$' & \mulc{3}{c}{Efficiency [\%]}  \\
         &      & Bkg.      &           &          & $\rm ee\ell\ell$&           &   $\rm eeqq$     & \mulc{3}{c}{}   \\
         &      &           &           &          &                 &           &                  &   ee   & $\mu\mu$ & $\tau\tau$ \\ \hline
    (3.1)  & 40948& 40899.7   &   7422.7  &   467.9  &    27011.1      &     260.1 &   5738.0         & 34.1   &  36.2    &   33.3   \\
    (3.2)  &  3203&  2816.0   &   1685.9  &   153.3  &      778.9      &      63.1 &    134.8         & 22.7   &  24.0    &   21.1    \\
    (3.3)  &  2031&  1912.0   &   1557.9  &   100.5  &      199.4      &      44.4 &      9.8         & 22.7   &  24.0    &   20.0    \\
    (3.4)  &  1359&  1247.1   &    939.8  &    83.2  &      182.2      &      32.5 &      9.3         & 21.8   &  23.4    &   19.5    \\
    (3.5)  &   572&   538.3   &    427.4  &    41.4  &       55.5      &      13.3 &      0.7         & 15.5   &  17.8    &   14.1    \\
    (3.6)  &   390&   361.8   &    273.4  &    29.9  &       52.5      &       5.8 &      0.2         & 14.7   &  17.3    &   12.6    \\\hline
    (3.7)  &    28&    22.3   &      4.4  &     4.0  &       13.3      &       0.5 &      0.1         & 14.6  &   17.2    &   11.9    \\
         &      & $\pm$1.6  & $\pm$0.7  &  $\pm$0.3&  $\pm$1.4       &   $\pm$0.1&   $\pm$0.0     &$\pm$2.0  &$\pm$2.0  &$\pm$2.1  \\ \hline
         &      &           &           &          &                 &           &                  & (41.0)   &  (48.8)    &  (33.4)   \\\hline 
\multicolumn{11}{c}{}   \\
\end{tabular}

 \label{t:sum}
  \centering
  \begin{tabular}{l|rr|rrrrr|rrr}
    \multicolumn{11}{l}{\bf Sum}\\
\hline
   ~~$\sum$~~  & ~~~~~55   & ~~~~44.4       & ~~~14.8   & ~~~6.8      &      ~~~~~21.8      &      ~~~1.0 &    ~~~~0.1       & 37.5    & 41.0     &   29.3   \\
         &      & $\pm$2.0  & $\pm$1.3  &  $\pm$0.3&  $\pm$1.5       &   $\pm$0.1&   $\pm$0.0     &$\pm$2.8  &$\pm$2.8  &$\pm$2.9  \\\hline
         &      &           &           &          &                 &           &                  & (105.6)   &  (116.1)    &  (82.5)   \\\hline 
             
  \end{tabular}
  \setcounter{table}{1}

  \caption[]{\label{t:results}
    \protect{\parbox[t]{12cm}{\sl
        The number of remaining events in the data after each cut,
        and the number expected from Standard Model background
        sources. Also shown are the efficiencies of expected signal events
        for a 130~GeV doubly-charged Higgs boson assuming ee, 
        $\mu\mu$ or $\tau\tau$ decays. The number of expected signal events
        for $\hee=0.1$ is shown in brackets assuming 100\% 
        branching ratio for the given decay mode. The errors due to Monte 
        Carlo statistics are also listed for events surviving the full analysis.
        }}
    }
\end{table}

%---------------------------------------------------------------------------

\subsection{Systematic Uncertainties}

\label{ss:syst}

The largest background in the selection is from processes with four charged 
leptons in the final state, particularly from multi-peripheral ``two-photon'' 
processes. Of concern is the fact that, in our standard Monte Carlo background
samples available at all centre-of-mass energies, the multi-peripheral 
diagrams are treated with specialised event generators which neglect 
interference with non-multi-peripheral diagrams.  Special samples of the full 
set of $\ee\ra\ee\ell^+\ell^-$ diagrams, including interference,
were prepared using grc4f2.2\cite{ref:grc4f} at $\sqrt{s}=206$~GeV to study 
this effect. The background using the full set of $\ee\ell^+\ell^-$ diagrams 
including interference is in both analyses about $25$\% lower than our standard 
set of Monte Carlo generators. While grc4f2.2 includes interference effects, 
it has other differences with respect to our standard background simulations 
and cannot be used as the primary sample. We therefore simply assign a 25\% 
systematic uncertainty on the $\ee\ell^+\ell^-$ background according to this 
cross-check.
Monte Carlo modelling of the variables used in the selection cuts can also 
induce systematic effects. The possible level of mismodelling is assessed by
comparing data and background Monte Carlo for each variable after the preselection 
(cut (2.1) and (3.1), respectively) where the contribution from a signal would 
be negligible. Differences between the data and background Monte Carlo simulation
are used to define a possible shift in each variable, and then the systematic 
uncertainties are evaluated by varying the cuts by these shifts.
Both the final expected background and signal efficiencies are re-calculated with 
these shifted cuts, and the full differences from the nominal values are assigned
as systematic uncertainties.

The uncertainty of charge identification, used in cuts~(2.7/3.7) in 
Section~\ref{ss:analysis} to reject a significant fraction of the background, is 
estimated from a clean sample of Bhabha events selected by changing the cuts as 
follows. The cuts (2.2)c and (3.2)d are not applied.
Cuts (2.2)d and (3.2)e are changed from $E_{\rm jet1}+E_{\rm jet2}(+E_{\rm jet3}) <  0.95 \sqrt{s}$
to $E_{\rm jet1}+E_{\rm jet2}(+E_{\rm jet3}) >  0.95 \sqrt{s}$.
This sample consists mainly of Bhabha events and has no overlap with the search 
sample. The fraction of like-sign electron pairs is 2.0\% in data and 1.7\% in 
Monte Carlo. The systematic uncertainties on the background and signal efficiencies 
are evaluated by randomly changing the sign of the charge for 0.15\% of the tracks, 
in order to increase the fraction of fake like-sign events by 0.3\%, the observed
difference between data and Monte Carlo in Bhabha sample.
The full differences between the new background and efficiencies and the nominal ones
are taken as systematic uncertainties.

The systematic uncertainties are summarised in Table~\ref{t:syst}. Additional 
systematic uncertainties, such as on the integrated luminosity, are negligible.

\begin{table}[ht]
\centering
  \begin{tabular}{l|c|c|c|c|c}
\hline
                           &                 &\multicolumn{2}{c|}{2-lepton analysis:} &\multicolumn{2}{c}{3-lepton analysis:}  \\
    Quantity               & Variation       & $\Delta$ Bkg & $\Delta$ Sig & $\Delta$ Bkg & $\Delta$ Sig\\
                           &                 &   (\%)       &   (\%)   &   (\%)       &   (\%)    \\\hline
   Jet $\cos{\theta}$      & $\pm0.5^\circ$  &    8         &    1     &     7        &    1      \\
  Jet Energy               &   $\pm$1\%      &    1         &    1     &     2        &    1      \\
   $\phi_{\rm acol}$       &  $\pm0.5^\circ$ &    2         &    1     &     1        &    1      \\
  Charge Misidentification &    0.15\%       &   14         &    1     &     4        &    1      \\\hline
  Background Modelling     & (see text)      &   25         &   --     &    25        &   --      \\
  Monte Carlo Statistics   &      --         &    8         &   10     &     7        &    14     \\\hline
 \mulc{2}{c|}{Quadratic Sum}                 &   31         &   10     &    27        &    14     \\\hline
  \end{tabular}
  \caption{\label{t:syst}
    \sl Systematic uncertainties on signal and background.}
\end{table}

%---------------------------------------------------------------------------

\subsection{Direct Search Results}

\begin{figure}[b]
  \begin{center}
      \mbox{\epsfig{file=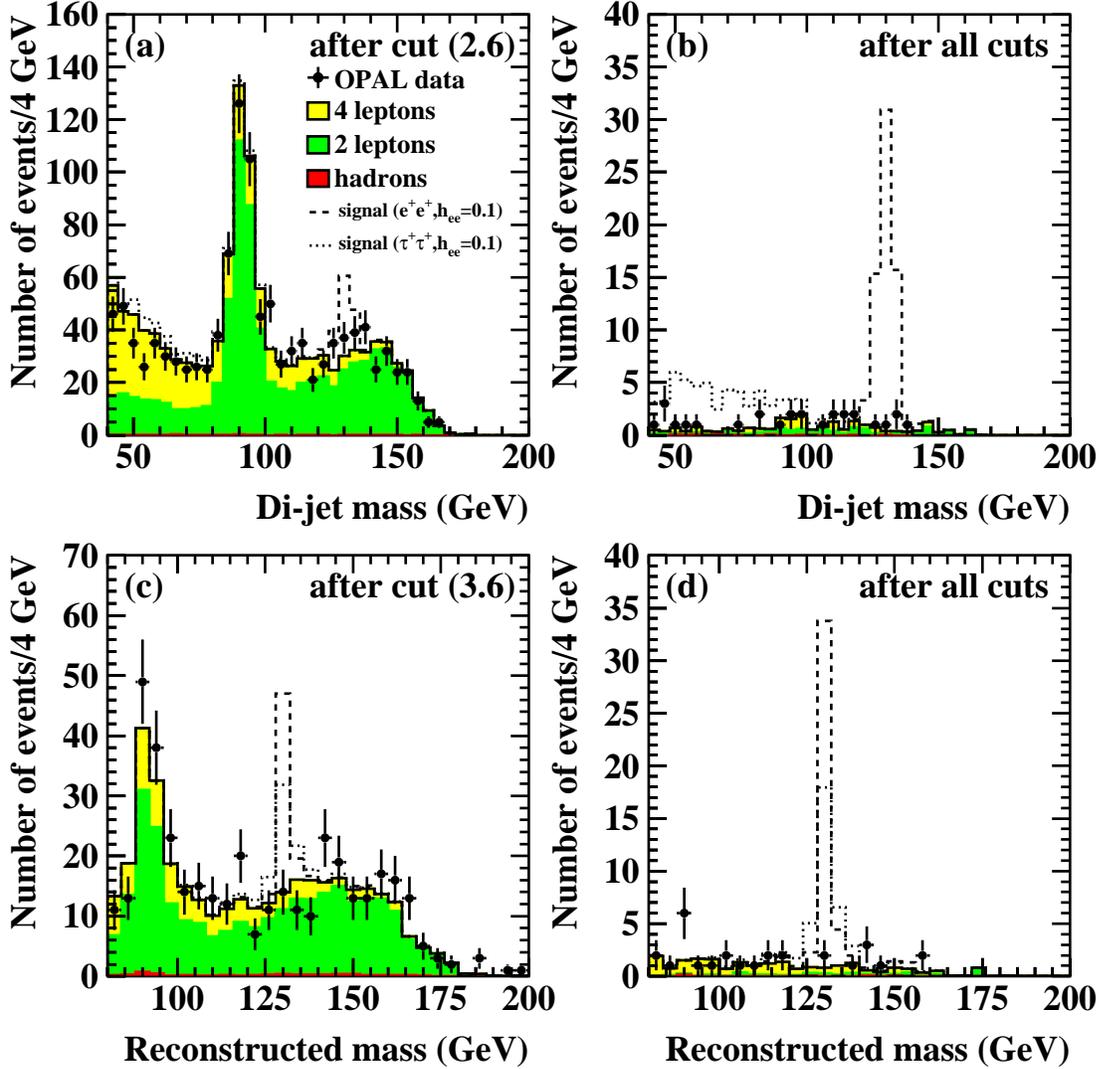,width=\textwidth
          }}
  \end{center}
\caption{\label{f:massinv}
  \protect{\sl
  The reconstructed $\hpmpm$ candidate mass distributions.
  The invariant di-jet mass is shown for the 2-lepton analysis 
  both before and after the like-signed jet requirement (cut (2.7)) in
  (a) and (b), respectively. For the 3-lepton analysis, the reconstructed
  $\hpmpm$ mass using the jet angles as discussed in the text, is shown
  before and after cut (3.7) in (c) and (d), respectively.
  The points with error bars indicate the OPAL data and  the shaded regions
  indicate the background expectation. Note that ``hadrons'' includes both
  $\qq (\gamma)$ and hadronic events from all 4-fermion processes.
  Two example signal expectations for a 130 GeV doubly-charged Higgs boson 
  are also shown normalised to a cross-section corresponding to  $\hee=0.1$,
  assuming either a 100\% $\hpmpm\ra\rm ee$ branching ratio (dashed line) 
  or a 100\% $\hpmpm\ra\tau\tau$ branching ratio (dotted line). Note that 
  due to the undetected neutrinos from the tau-lepton decay there is
  no peak in the $\hpmpm\ra\tau\tau$ signal sample of the 2-lepton analysis
  ((a) and (b)).
  }
}
\end{figure}

In the two-lepton analysis the invariant mass of the two jets is calculated 
using the measured jet energies and directions, because it is not possible 
to use the ``angle-based'' kinematic reconstruction described in 
section~\ref{ss:analysis} for the three-lepton analysis.
The mass distribution is shown in Figure~\ref{f:massinv} for events passing 
all cuts except the like-signed charge requirement~(a), and also with all 
cuts applied~(b). No excess of events which could imply the presence of a 
signal is observed in the data.

In the three-lepton analysis we calculate the $\hpmpm$ candidate reconstructed 
masses, $M_{\rm rec}$,shown in Figure~\ref{f:massinv}, using the ``angle-based'' 
kinematic reconstruction described in item~(3.3) in Section~\ref{ss:analysis}.
The mass distributions are shown both for events passing all cuts except the 
like-signed charge requirement~(c), and also with all cuts applied~(d). 
Additionally, as a cross-check to ensure that no di-jet mass peak present after 
the event reconstruction is reduced by the angle-based method, the largest 
di-jet mass calculated from only the track and cluster information 
(Section~\ref{ss:analysis}) was examined. 
No excess of events which could imply the presence of a signal is observed in 
the data.

Limits are set on the $\hpmpm$ Yukawa coupling $\hee$, assuming that the sum of 
the branching fractions of the $\hpmpm$ to all lepton flavour combinations is 
100\%. The efficiency for an arbitrary Higgs boson mass is determined by linear 
interpolation between the simulated signal Monte Carlo samples.
The number of observed events, together with the number of expected signal and 
background events from both the two-lepton and three-lepton analyses are combined 
using the likelihood ratio method described in \cite{ref:tom}, which incorporates
the systematic uncertainties into the limits using a numerical convolution technique.
For the purpose of extracting the limits, a $\pm 10$~GeV ``sliding mass window'' 
around the hypothetical Higgs boson mass is used. Events within this window are 
counted in data and Monte Carlo simulation. The hypothetical Higgs boson mass is
varied in 1~GeV steps. The width of the mass window is chosen such that it contains
most of the expected signal events. A small efficiency correction, typically around 
5\% for ee and $\mu\mu$ and 10\% for $\tau\tau$, due to this window is applied.
In the two-lepton analysis for any channel containing $\tau$ leptons no mass window 
cut is applied, because in this channel it is not possible to reconstruct the 
correct mass of the doubly-charged Higgs boson due to the undetected neutrinos.

\begin{figure}[b]
  \begin{center}
      \mbox{\epsfig{file=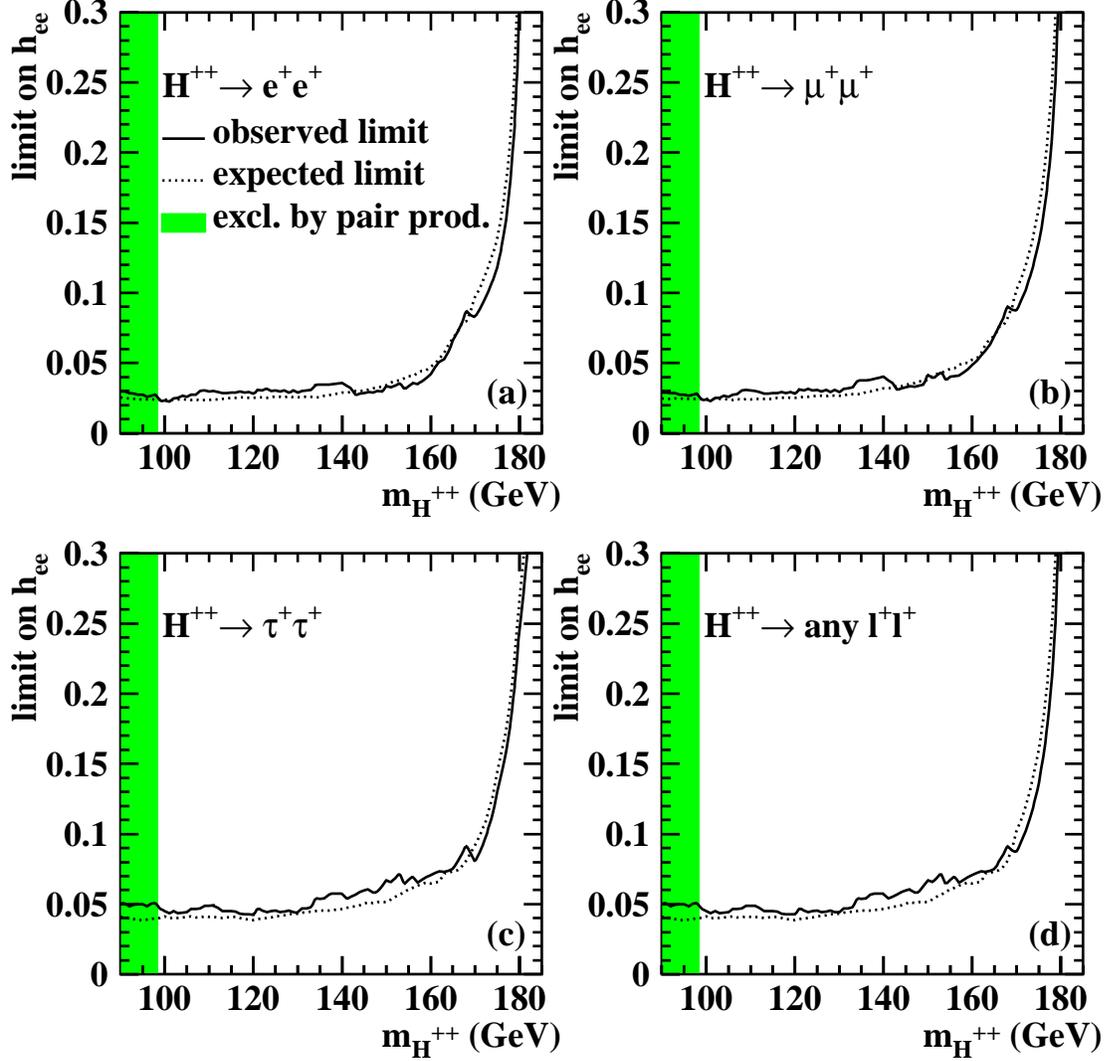,width=\textwidth
          }}
  \end{center}
\caption{\label{f:limyukawa}
  \protect{\sl
  Limits at the 95\% confidence level on the Yukawa coupling $\hee$
  assuming a 100\% branching fraction of the $\hpmpm$ to (a) ee, (b) 
  $\mu\mu$ and (c) $\tau\tau$. The limits are calculated with the
  combined results of the two-lepton and three-lepton analysis.
  In (b) and (c), the limits should be regarded as valid in the large 
  branching fraction limit, since non-zero $\hee$ implies a non-zero 
  electron branching fraction (see text).
  Since the ee and $\mu\mu$ efficiencies and mass resolutions are 
  very similar, figures (a) and (b) are almost identical.
  The median expected limits assuming only Standard Model processes 
  are shown by the dotted lines, while the actual limits inferred 
  from the data are shown by the solid lines.
  In figure (d) the limit for arbitrary lepton flavour combinations
  (ee, e$\mu$, e$\tau$, $\mu\mu$, $\mu\tau$ and $\tau\tau$) is shown.
  It is determined by the pure $\tau\tau$ case except for masses in
  excess of 170~GeV.
  The shaded regions for masses below 98.5~GeV are excluded
  in Left-Right symmetric models by the OPAL pair production search 
  \cite{ref:pr349}.
  }
}
\end{figure}

The limits on $\hee$ are calculated using the efficiencies determined from
the PYTHIA Monte Carlo samples and the production cross-sections are determined in 
a consistent manner using PYTHIA (see discussion in Section~\ref{ss:data}).
No systematic uncertainty is assigned for theoretical uncertainties.
The 95\% confidence level limits on $\hee$ from combining both analyses are
shown in Figure~\ref{f:limyukawa}(a)--(c) assuming a branching fraction
of the doubly-charged Higgs boson into ee, $\mu\mu$, $\tau\tau$ of
100\%, respectively. Strictly, due to the production mechanism involving
non-zero $\hee$, exactly 100\% $\mu\mu$ or $\tau\tau$ decays are not
possible, therefore the latter limits should be considered for the case
$\mathrm{h}_{\mu\mu,\tau\tau} \gg \hee$.
In Figure~\ref{f:limyukawa}(d), for each mass the highest limit from all possible 
lepton flavour combinations is shown. An upper limit on $\hee<$~0.071 is inferred for
$M(\hpmpm)<$~160~GeV at the 95\% confidence level, which is valid for all possible 
lepton flavour combinations in the decays. The limit is determined by the pure 
$\tau\tau$ case except for masses in excess of 170~GeV.
For the case of pure \rm{ee} decays the limit is $\hee<$~0.042, and for $\mu\mu$ 
decays $\hee<$~0.049, both for $M(\hpmpm)<$~160~GeV. For the mixed flavour decay
modes e$\mu$, e$\tau$, and $\mu\tau$ the limit is between those for pure decays of 
the two involved flavours.

%=======================================================================

\section{Indirect Search}

\label{s:indirect}

Doubly-charged Higgs bosons would contribute to Bhabha scattering via $t$-channel 
exchange as shown in Figure~\ref{f:indirect}. The Born level differential 
cross-section for Bhabha scattering including the exchange of a doubly-charged Higgs 
boson with right-handed couplings has been calculated in~\cite{ref:swartz}. 
At high masses, $M(\hpmpm) \gg \sqrt{s}$, the cross-section is identical to that 
derived for four-fermion contact interactions with right-handed currents~\cite{ref:Eichten}
($\eta_{\rm RR}=1$, $\eta_{\rm LL} = \eta_{\rm LR}=0$), with the replacement
of $g/\Lambda$ by $\hee/M(\hpmpm)$ where $\hee$ is the Higgs coupling
to electrons\footnote{In~\cite{ref:swartz} $\hee$ is denoted $g_{\rm ee}$.}. 
At values of $M(\hpmpm)$ comparable to the centre-of-mass energy, this correspondence 
is modified by the inclusion of a propagator term. For comparison with the 
experimental data, QED radiative corrections are applied to the Born level terms for
doubly-charged Higgs boson exchange and interference with Standard Model
processes given in~\cite{ref:swartz} using the program MIBA~\cite{ref:MIBA}. 
Initial state radiation is calculated up to ${\cal O}(\alpha^2)$ in the 
leading log approximation with soft photon exponentiation, and the 
${\cal O}(\alpha)$ leading log final state QED correction is applied.
The BHWIDE~\cite{ref:BHWIDE} program is used to calculate the Standard Model 
contribution to the differential cross-section. The theoretical predictions are 
calculated using the same acceptance cuts as are applied to the data. 

This analysis uses OPAL measurements of the differential cross-section for 
$\ee\rightarrow\ee$ at centre-of-mass energies of 183--209~GeV\cite{ref:OPAL-SM183, ref:opal2f}. 
The data between 203~GeV and 209~GeV are grouped into two sets with mean energies 
of approximately 205~GeV and 207~GeV.
The total integrated luminosity of the data amounts to
688.4~pb$^{-1}$. These measurements cover the range $|\cos \theta | < 0.9$, in 15 bins of
$\cos \theta$ (as defined in~\cite{ref:opal2f}), and correspond 
to $\theta_{\mathrm{acol}} < 10^\circ$ where $\theta_{\mathrm{acol}}$ is the acollinearity 
angle between electron and positron. 
It is verified that the effect of doubly-charged Higgs boson exchange on the low-angle 
Bhabha scattering cross-section has a negligible effect on the luminosity determination 
even for values of $\hee$ a few times larger than excluded by this measurement.

\begin{figure}[b]
  \begin{center}
      \mbox{\epsfig{file=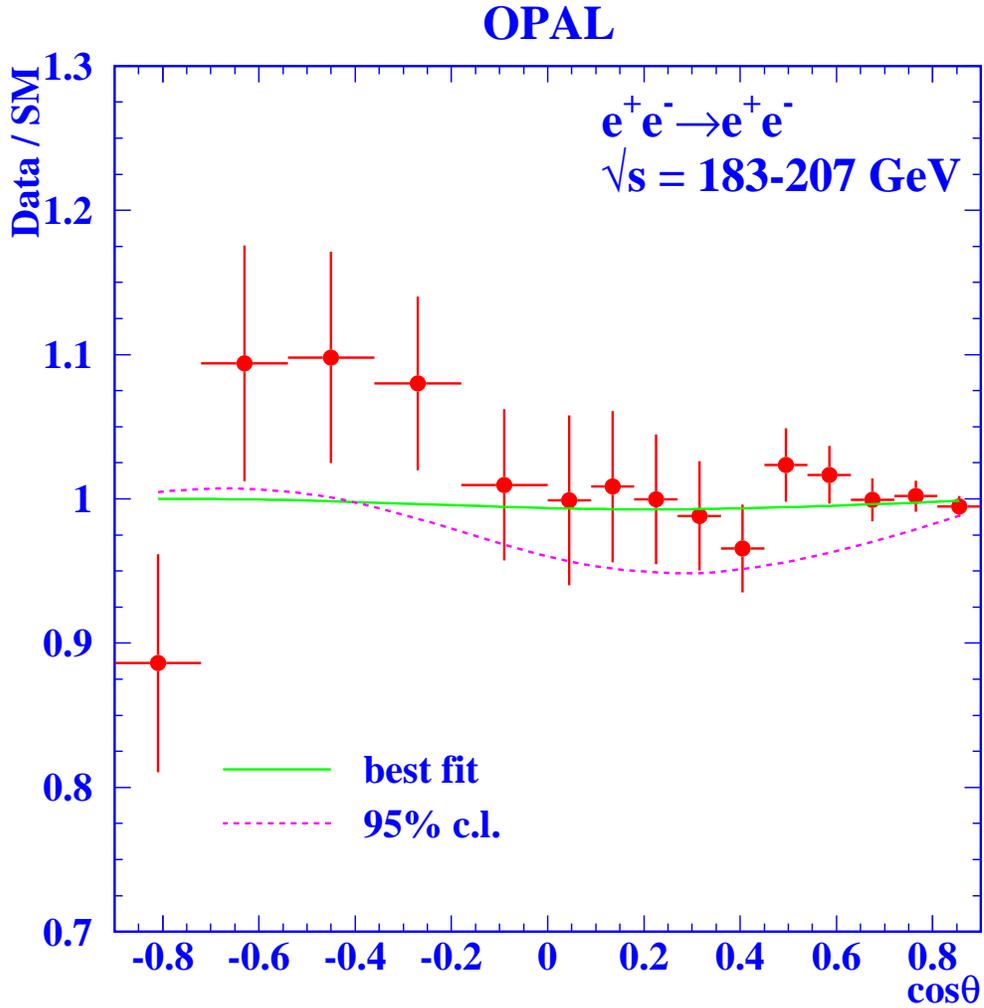,width=0.9\textwidth
          }}
  \end{center}
\caption{\label{f:ee207}
  \protect{\sl
  Ratio of the measured luminosity-weighted average differential 
  cross-section for $\ee\rightarrow\ee$ at 183--207~GeV to the 
  Standard Model prediction.
  The points with error bars show the OPAL data, while the curves 
  show theoretical predictions for a doubly-charged Higgs boson mass 
  of 130 GeV. The solid curve corresponds to the best fit to all 
  data, the dashed curve corresponds to a coupling equal to the 95\% 
  confidence level limit.
  }
}
\end{figure}

\begin{figure}[b]
  \begin{center}
      \mbox{\epsfig{file=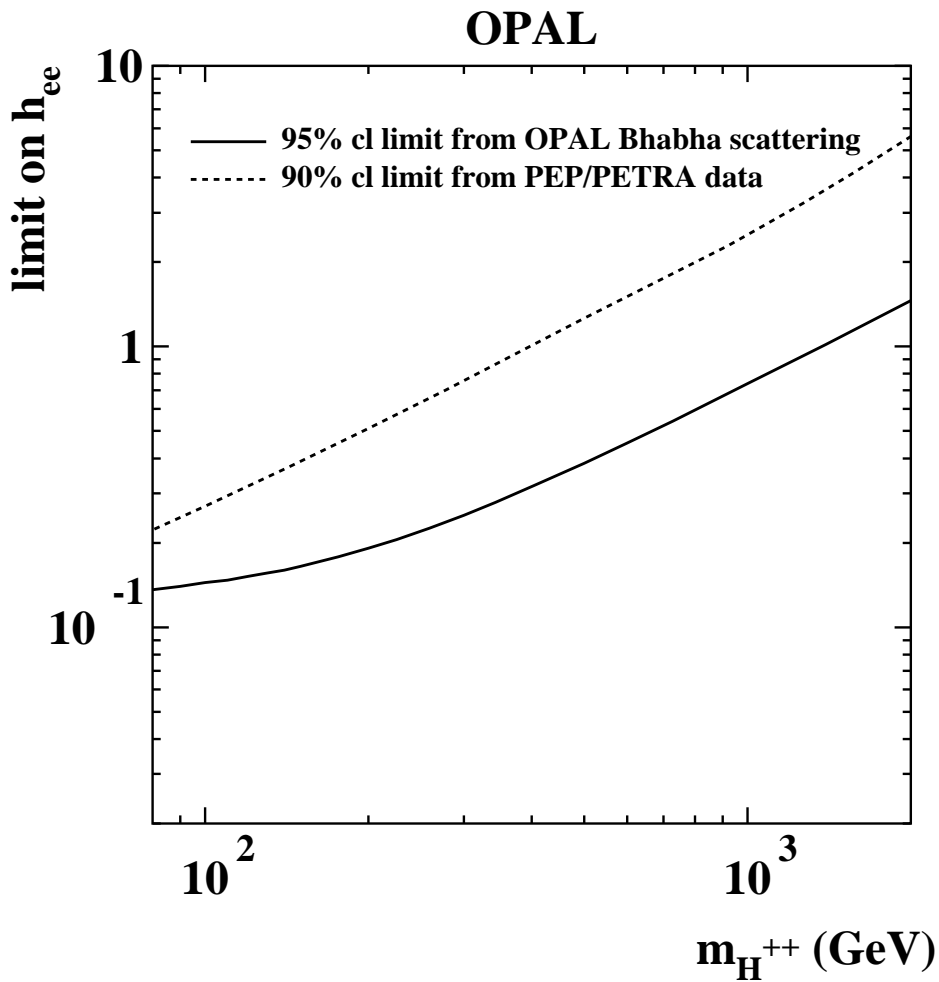,width=0.9\textwidth
          }}
  \end{center}
\caption{\label{f:limindir}
  \protect{\sl
  Limits at the 95\% confidence level on the Yukawa coupling $\hee$ 
  as a function of $M(\hpmpm)$ derived from Bhabha scattering data 
  (solid line) for an $\hpmpm$ coupling to right-handed particles.
  Limits at 90\% confidence level derived from \mbox{PEP} and 
  \mbox{PETRA} data~\cite{ref:swartz} are shown, as a dashed line, 
  for comparison.
  }
}
\end{figure}

\begin{figure}[b]
  \begin{center}
      \mbox{\epsfig{file=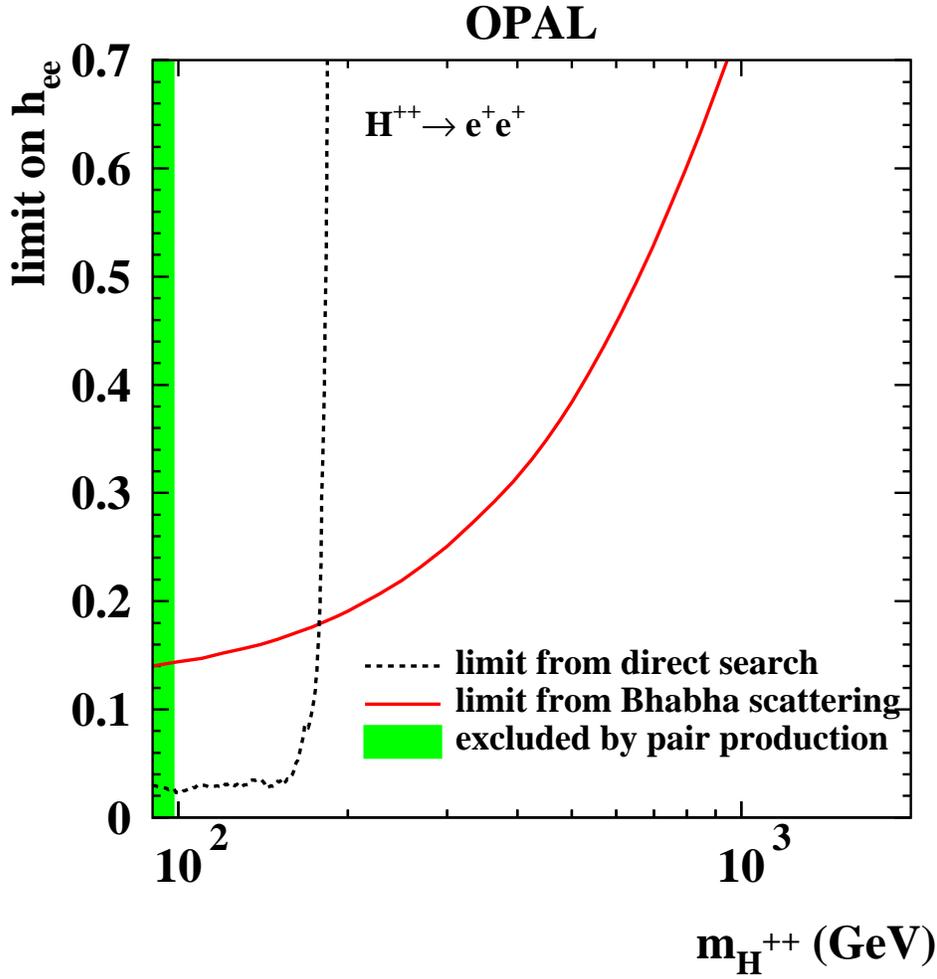,width=0.9\textwidth
          }}
  \end{center}
\caption{\label{f:limyukawabhabha}
  \protect{\sl 
  Limits at the 95\% confidence level on the Yukawa coupling $\hee$
  assuming a 100\% branching fraction of the $\hpmpm\ra ee$. The 
  direct limit is calculated with the combined results of the 
  two-lepton and three-lepton analyses.
  The indirect limit on $\hee$ obtained from Bhabha scattering
  described in Section~\ref{s:indirect} is also shown.
  The shaded regions for masses below 98.5~GeV are excluded
  in  Left-Right Symmetric models  by the OPAL pair production
  search \cite{ref:pr349}.
  }
}
\end{figure}

The measured differential cross-sections are fitted with the theoretical prediction 
using a $\chi^2$ fit. The fit is performed for fixed values of the doubly-charged Higgs 
boson mass between 80~GeV and 2000~GeV, allowing the square of the coupling, $\heesq$, 
to vary. Although only $\heesq >0$ is physically meaningful, in order to allow for 
the case where the data fluctuate in the opposite direction to that expected for 
doubly-charged Higgs boson exchange, both positive and negative values of $\heesq$ are 
allowed in the fit. Experimental and theoretical systematic uncertainties and 
their correlations are treated as discussed in~\cite{ref:opal2f}. 
The fitted values of $\heesq$ are consistent with zero for all masses,
indicating that the data are consistent with the Standard Model prediction.
For example, for a mass of 130~GeV the fitted value of $\heesq$ is
0.003$\pm$0.011, and the fit has a $\chi^2$ of 97.0 for 119 degrees of
freedom. Figure~\ref{f:ee207} shows the ratio of the measured
luminosity-weighted average differential cross-section at 183--207~GeV 
to the Standard Model prediction, together with the results of the 
fit. 95\% confidence level limits on the coupling as a function of mass
were derived by integrating the likelihood function obtained from $\chi^2$
over the region $\heesq >0$, and are shown in Figure~\ref{f:limindir}. 
The limits are considerably more stringent than those derived from PEP 
and PETRA data~\cite{ref:swartz}. Figure~\ref{f:limyukawabhabha} shows the limits
from the indirect search together with those from the direct search. The
indirect limits are less restrictive than those from the direct search 
at low masses, but extend to much higher masses.

%=======================================================================

\section{Conclusion}
\label{s:conclusion}

   A direct search for the single production of doubly-charged Higgs bosons
   has been performed.
   No evidence for the existence of $\hpmpm$ is observed.
   Upper limits are determined on the Higgs Yukawa coupling to like-signed
   electron pairs, $\hee$.
   A 95\% confidence level upper limit of $\hee<$~0.071 is inferred for 
   $M(\hpmpm)<$~160~GeV assuming that the sum of the branching fractions of
   the $\hpmpm$ to all lepton flavour combinations is 100\%.
   Additionally, indirect constraints on $\hee$ for $M(\hpmpm)<$~2~TeV
   are derived from Bhabha scattering where the $\hpmpm$ would contribute 
   via $t$-channel exchange for $M(\hpmpm)<$~2~TeV.
   These are the first results on both the single production search
   and constraints from Bhabha scattering reported from LEP.

%=======================================================================
\bigskip

{\bf\Large Acknowledgements}

\bigskip

The authors would like to thank Andr\'e Sch\"oning for suggesting that we 
perform this search, Steve Godfrey and Pat Kalyniak for valuable discussions
and assistance during the preparation of this paper, and also Emmanuelle Perez 
for a helpful hint about the PYTHIA code.

We particularly wish to thank the SL Division for the efficient operation
of the LEP accelerator at all energies
 and for their close cooperation with
our experimental group.  In addition to the support staff at our own
institutions we are pleased to acknowledge the  \\
Department of Energy, USA, \\
National Science Foundation, USA, \\
Particle Physics and Astronomy Research Council, UK, \\
Natural Sciences and Engineering Research Council, Canada, \\
Israel Science Foundation, administered by the Israel
Academy of Science and Humanities, \\
Benoziyo Center for High Energy Physics,\\
Japanese Ministry of Education, Culture, Sports, Science and
Technology (MEXT) and a grant under the MEXT International
Science Research Program,\\
Japanese Society for the Promotion of Science (JSPS),\\
German Israeli Bi-national Science Foundation (GIF), \\
Bundesministerium f\"ur Bildung und Forschung, Germany, \\
National Research Council of Canada, \\
Hungarian Foundation for Scientific Research, OTKA T-038240, 
and T-042864,\\
The NWO/NATO Fund for Scientific Research, the Netherlands.\\

%=======================================================================

%=======================================================================

\newpage
\clearpage

\newpage
\clearpage

\newpage
\clearpage

\newpage
\clearpage

\newpage
\newpage
\clearpage

\newpage

%=======================================================================

\end{document}